\documentclass[aip,apl,twocolumn,superscriptaddress,citeautoscript,reprint]{revtex4-2}

\usepackage{blindtext}
\usepackage{graphicx}
\usepackage{bm}
\usepackage{epsfig}
\usepackage{amssymb}
\usepackage{amsfonts}
\usepackage{braket}
\usepackage{color}
\usepackage{epstopdf}
\usepackage{hyperref}
\usepackage{float}
\restylefloat{table}
\usepackage{bibentry}
\usepackage{color}
\usepackage{multirow}
\usepackage[caption=false]{subfig}
\usepackage{ulem}

\usepackage{amsfonts}
\usepackage{amsthm}
\usepackage{graphicx}
\usepackage{multirow}
\usepackage{color}
\usepackage{bbold}
\usepackage{bm}
\usepackage{times}
\usepackage{amsmath,bm,amsfonts}
\usepackage{dcolumn}
\usepackage{graphicx}
\usepackage{latexsym}

\newcommand{\bpm}{\begin{pmatrix}}
	\newcommand{\epm}{\end{pmatrix}}
\newcommand{\ba}{\begin{eqnarray}}
	\newcommand{\ea}{\end{eqnarray}}
\newcommand{\bd}{\begin{displaymath}}

	\graphicspath{{figures/}}
	
\begin{document}

\title{Capping and gate control of anomalous Hall effect and hump structure in ultra-thin SrRuO$_3$ films}

\author{Donghan Kim}
\thanks {These authors contributed equally to this work.}
\affiliation{Center for Correlated Electron Systems, Institute for Basic Science, Seoul 08826, Korea}
\affiliation{Department of Physics and Astronomy, Seoul National University, Seoul 08826, Korea}

\author{Byungmin Sohn}
\thanks {These authors contributed equally to this work.}
\affiliation{Center for Correlated Electron Systems, Institute for Basic Science, Seoul 08826, Korea}
\affiliation{Department of Physics and Astronomy, Seoul National University, Seoul 08826, Korea}

\author{Minsoo Kim}
\affiliation{Center for Correlated Electron Systems, Institute for Basic Science, Seoul 08826, Korea}
\affiliation{Department of Physics and Astronomy, Seoul National University, Seoul 08826, Korea}

\author{Sungsoo Hahn}
\affiliation{Center for Correlated Electron Systems, Institute for Basic Science, Seoul 08826, Korea}
\affiliation{Department of Physics and Astronomy, Seoul National University, Seoul 08826, Korea}

\author{Youngdo Kim}
\affiliation{Center for Correlated Electron Systems, Institute for Basic Science, Seoul 08826, Korea}
\affiliation{Department of Physics and Astronomy, Seoul National University, Seoul 08826, Korea}

\author{Jong Hyuk Kim}
\affiliation{Department of Physics, Yonsei University, Seoul 03722, Korea}

\author{Young Jai Choi}
\affiliation{Department of Physics, Yonsei University, Seoul 03722, Korea}

\author{Changyoung Kim}
\email[Electronic address:$~~$]{changyoung@snu.ac.kr}
\affiliation{Center for Correlated Electron Systems, Institute for Basic Science, Seoul 08826, Korea}
\affiliation{Department of Physics and Astronomy, Seoul National University, Seoul 08826, Korea}

\date{\today}

\begin{abstract}
Ferromagnetism and exotic topological structures in SrRuO$_3$ (SRO) induce sign-changing anomalous Hall effect (AHE). Recently, hump structures have been reported in the Hall resistivity of SRO thin films, especially in the ultra-thin regime. We investigate the AHE and hump structure in the Hall resistivity of SRO ultra-thin films with an SrTiO$_3$ (STO) capping layer and ionic liquid gating. STO capping results in sign changes in the AHE and modulation of the hump structure. In particular, the hump structure in the Hall resistivity is strongly modulated and even vanishes in STO-capped 4 unit cell (uc) films. In addition, the conductivity of STO-capped SRO ultra-thin films is greatly enhanced with restored ferromagnetism. We also performed ionic liquid gating to modulate the electric field at SRO/STO interface. Drastic changes in the AHE and hump structure are observed with different gate voltages. Our study shows that the hump structure as well as the AHE can be controlled by tuning inversion symmetry and the electric field at the interface.
\end{abstract}
\maketitle

Ultra-thin film systems can host a range of physical properties that differ from those of bulk systems due not only to the reduced dimension but also to the natural inversion symmetry breaking (ISB)~\cite{fowlie2017conductivity,xia2009critical}. Thin films of an itinerant ferromagnetic oxide, SrRuO$_3$ (SRO), have recently attracted widespread attention because of unresolved phenomena in the ultra-thin limit. For example, the anomalous Hall effect (AHE) becomes very small, and even reverses its sign, in ultra-thin SRO films due to the change in the Berry phase in the momentum space~\cite{haham2011scaling,sohn2019sign,groenendijk2020berry}. 

Another intriguing phenomenon, a hump structure in the Hall resistivity, also emerges in the ultra-thin limit~\cite{groenendijk2020berry,sohn2018emergence,qin2019emergence,gu2019interfacial,kan2018alternative,wang2020controllable,kimbell2020two,wu2018artificial,wu2020berry,wysocki2020electronic,wysocki2020validity}. Various models have been suggested to explain the origin of the hump structure. There are several recent reports explaining the origin of the hump structure by the superposition of different AHEs and coercive fields from inhomogeneous regions~\cite{kan2018alternative,wang2020controllable,kimbell2020two,wu2018artificial,wu2020berry,wysocki2020electronic,wysocki2020validity}. Another proposal is based on the topological Hall effect (THE) model caused by non-coplanar magnetic structures such as magnetic skyrmions. In this conjecture, the finite Dzyaloshinskii-Moriya interaction (DMI) required for such magnetic structures appears
 due to distortions (atomic rumpling) near the surface~\cite{sohn2018emergence,qin2019emergence,gu2019interfacial} which are induced by the inversion symmetry breaking (ISB) at the surface~\cite{fowlie2017conductivity,kumah2014tuning,kumah2014effect,aso2013atomic,kan2015research}. This issue is under debate and being actively studied. 

On the other hand, independent of the origin, capability to control the features in the Hall effect should be an important subject. As one of the models suggests that ISB plays a key role in the emergence of the hump structure, ISB (i.e., electric field) may be used as a tuning parameter to manipulate the hump structure (and AHE). If the hump structure can be indeed controlled by tuning the ISB or an electric field, it could be useful for new device applications. It was previously demonstrated that inversion symmetry can be recovered with a capping layer~\cite{groenendijk2020berry, wang2019spin}. In this work, we study the AHE and hump structure in the Hall resistivity of SRO ultra-thin films for various thicknesses of STO capping layer. We also performed ionic liquid gating experiments to control the inversion symmetry. Through our capping and ionic liquid gating experiments, we demonstrate that AHE and hump structure are sensitive to a modest perturbation and thus can be controlled.

\begin{figure*}[htbp]
	\includegraphics[width=0.75\textwidth]{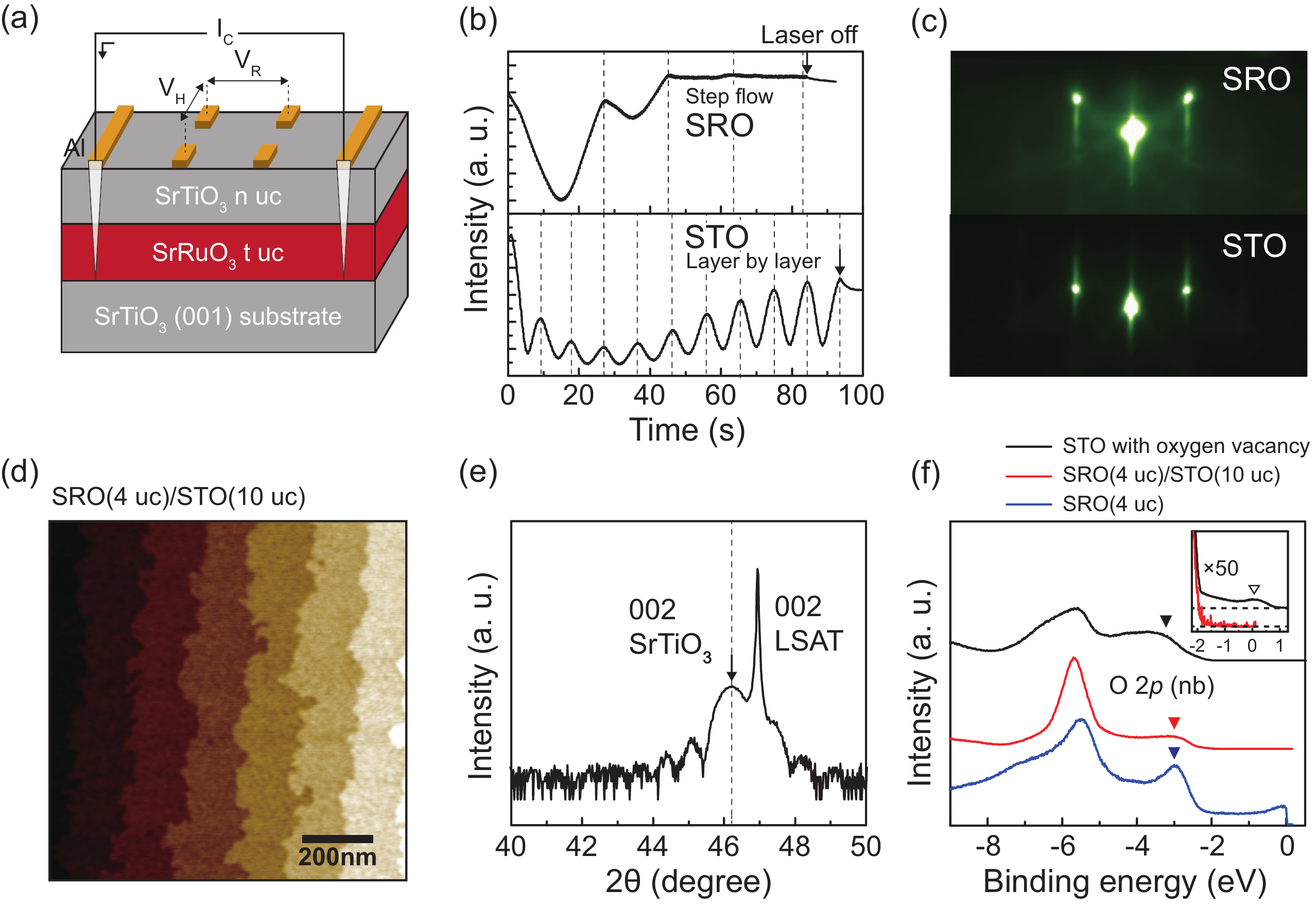}
	\caption{{\bf Characterization of films.} ({\bf a}) Schematic of SrRuO$_3$ (SRO) / SrTiO$_3$ (STO) thin films on an STO (001) substrate. {\it t} unit cell (uc) ({\it t} = 2 $\sim$ 4) SRO is capped with {\it n}~uc ({\it n} = 1 $\sim$ 10) STO. ({\bf b}) {\it In-situ} reflection high-energy electron diffraction (RHEED) intensity from (upper) 4~uc SRO and (lower) 10~uc STO heterostructures. ({\bf c}) RHEED patterns of SRO and STO layers. ({\bf d}) Surface topography of a 4~uc SRO thin film capped with 10~uc STO. ({\bf e}) X-ray diffraction $\theta$-2$\theta$ scan of a 50~uc STO thin film on a (LaAlO$_3$)$_{0.3}$(Sr$_2$TaAlO$_6$)$_{0.7}$ (LSAT) (001) substrate. The position of the STO (002) peak is consistent with the report in Ref.~\cite{haislmaier2016stoichiometry}, which shows that our STO thin film is stoichiometric. ({\bf f}) Photoemission spectroscopy data of 4~uc SRO (blue), 10~uc STO on 4~uc SRO (red), and STO with oxygen vacancies (black). Filled triangles mark the O 2{\it p} peak. The position of the O 2{\it p} peak of STO on 4~uc SRO is similar to that of 4~uc SRO. For the data from a deoxidized STO substrate (black), the O 2{\it p} peak appears at a higher binding energy due to the insulating nature of the substrate: the spectrum in the figure has been shifted by +1.3~eV for comparison. In-gap states (open triangle in the inset) emerge in the deoxidized STO, whereas the STO on 4~uc SRO does not show in-gap states (inset).}
	\label{fig:1}
\end{figure*}

High-quality STO-capped ultra-thin SRO films were grown on atomically flat TiO$_2$-terminated STO (001) substrates (Shinkosha Co. Ltd.) by pulsed laser deposition (PLD). The STO substrates were pre-annealed at 1,100~$^{\circ}{\rm C}$ under an oxygen partial pressure of $5\times 10^{-6}$~Torr. SRO layers were deposited at 700~$^{\circ}{\rm C}$ with an oxygen partial pressure of 100~mTorr. A KrF excimer laser was used with an energy fluence of 2~${\rm J/cm^2}$ and a 2~Hz repetition rate. After the SRO growth, STO capping layers were deposited at 700~$^{\circ}{\rm C}$ with an oxygen partial pressure of 10~mTorr, energy fluence of 1.2~${\rm J/cm^2}$ and 2~Hz repetition rate. The entire growth process was monitored using {\it in-situ} reflection high energy electron diffraction (RHEED).  

For transport measurements, 60~nm Au electrodes were deposited on the films using an e-beam evaporator. Ohmic contacts to the STO and SRO layers were made by ultrasonic wire bonding with Al wires~\cite{inoue2015origin,swartz2014spin,gunkel2016defect}. Transport measurements were conducted using a Physical Property Measurement System (PPMS; Quantum Design Inc.). The current-voltage characteristics of our heterostructure films confirm that the contact is ohmic (see Supplementary Material for details). Resistivity and Hall measurements were conducted by measuring V$_{\rm R}$ and V$_{\rm H}$ with the four-probe method. The resistivity of STO is about ${10^{16}}$ times larger than that of SRO~\cite{matsumoto2007electrical} (see Supplementary Material for details). Thus, we only considered the thickness of SRO in the calculation of the resistivity of SRO/STO heterostructures.
Diethylmethyl(2-methoxyethyl)ammonium bis(trifluoromethylsulfonyl)imide [DEME-TFSI] was used as the ionic liquid for ionic liquid gating. An Au top electrode was used for application of the gate voltage, which was set at 220~K.

Figure 1(a) shows a schematic diagram of a {\it t} unit cell (uc) SRO thin film on an STO substrate capped with {\it n}~uc STO. Figures 1(b) and (c) show the RHEED intensity plots and patterns for the SRO and STO. The deposition rate of the SRO (STO) was 0.05 (0.1)~uc/s. Figure 1(d) shows an atomic force microscopy (AFM) topographic image of a 4~uc SRO film with a 10~uc STO capping layer, and demonstrates the atomically flat surface of the SRO/STO heterostructures.  

To optimize the growth conditions for the STO thin films, the stoichiometry of the STO thin films is verified by X-ray diffraction (XRD) and {\it in-situ} ultraviolet photoelectron spectroscopy. It was suggested in a previous report that the stoichiometry of the STO thin film can be verified by measuring the out-of-plane $c$-axis lattice constant~\cite{haislmaier2016stoichiometry}. We grew 50~uc STO thin films on a (LaAlO$_3$)$_{0.3}$(Sr$_2$TaAlO$_6$)$_{0.7}$ (LSAT) substrate and performed $\theta$-2$\theta$ measurements (Fig. 1(e)). We found that the $c$-axis lattice constant of our STO film, $c =3.93~\AA$, is consistent with the previous result~\cite{haislmaier2016stoichiometry}, showing that the ratio between the Sr and Ti atoms is stoichiometric.  

Figure 1(f) shows He-I$\alpha$ (21.2~eV) photoemission spectra from a 4~uc SRO (blue), a 10~uc STO capping layer on 4~uc SRO (red), and a deoxidized STO substrate (black).  In-gap states are observed in the deoxidized STO substrate spectrum, whereas in-gap states do not appear in the spectrum from STO on 4 uc SRO film (inset of Fig. 1(f)). A previous study reported that the amount of oxygen deficiencies can be estimated by the intensity of in-gap states~\cite{pal2014chemical}. As the in-gap state spectral weight of our STO-capped film is smaller than that of STO with the least oxygen deficiency ($\delta = 0.004$) in the report~\cite{pal2014chemical}, we conclude that our STO capping layer has, if any, a negligible amount of oxygen vacancy. 

\begin{figure}[htbp]
	\includegraphics[width=0.46\textwidth]{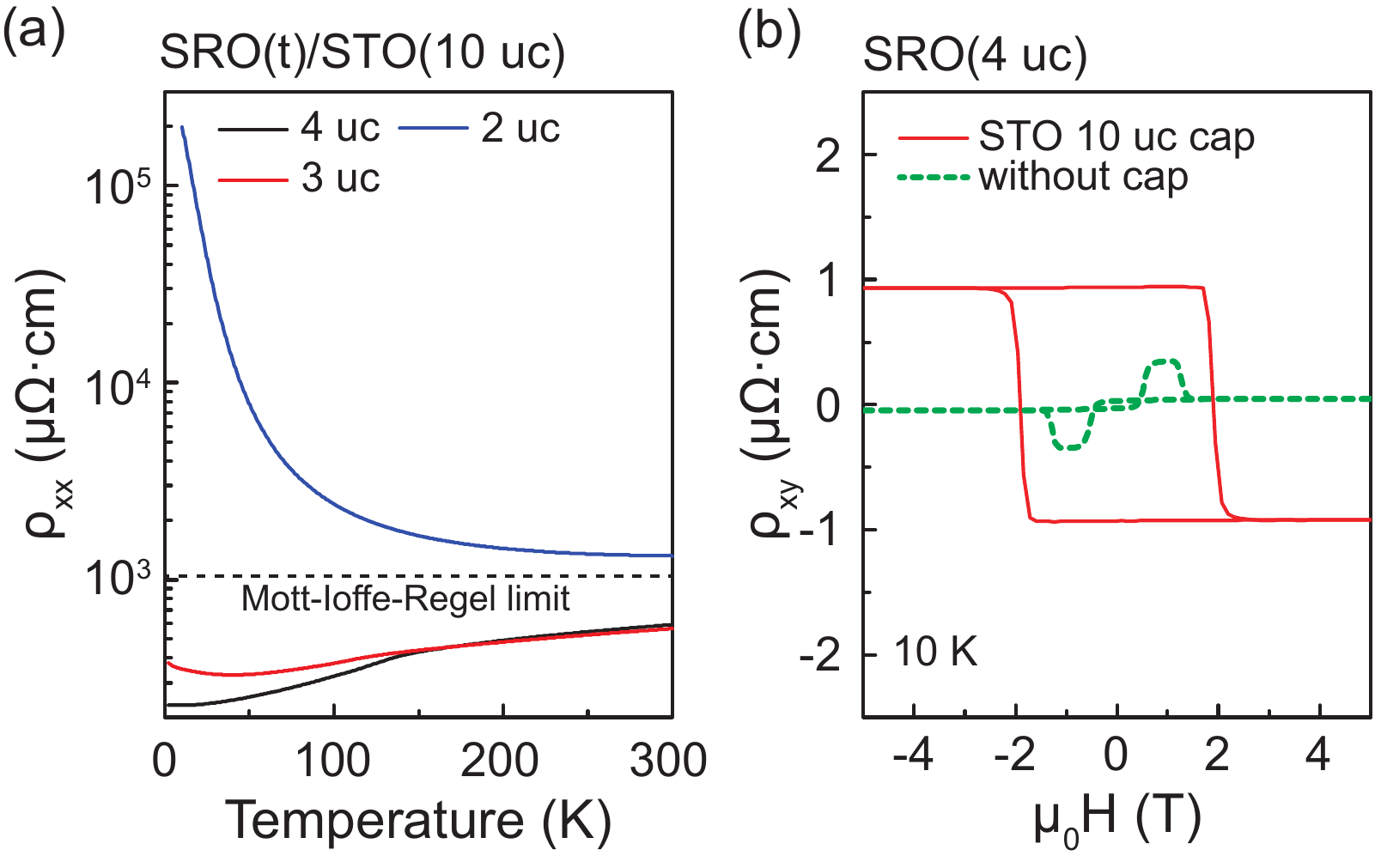}
	\caption{{\bf Resistivity and Hall effect results of {\it t}~uc ({\it t} = 2 $\sim$ 4) SRO films with 10~uc STO capping layer.} ({\bf a}) Resistivity of 10~uc STO-capped SRO for a range of thicknesses. A metal-insulator transition occurs between 2 and 3~uc SRO films. ({\bf b}) Hall effect measurements of 4~uc SRO thin films with and without the STO capping layer. The sign of the anomalous Hall effect (AHE) changes and the hump structure almost disappears when STO is capped on the 4~uc SRO film.}
	\label{fig:2}
\end{figure}

We measured the resistivity of SRO/STO heterostructures with various SRO thicknesses. Figure 2(a) shows the thickness-dependent resistivity of the SRO thin films with a 10~uc STO capping layer. The Mott-Ioffe-Regel (MIR) limit is used to identify the threshold between the metal and insulating states of SRO film~\cite{kang2019orbital,gunnarsson2003colloquium}. The MIR limit (represented by the horizontal dashed line in Fig. 2(a)), at which the mean free path becomes comparable to the interatomic distance, is given as $\rho_{\rm MIR} = ha/e^2$, where $h$, $a=3.93~\AA$ and $e$ are the Planck constant, lattice constant and electron charge, respectively~\cite{hussey2004universality,werman2017non}. Here the STO-capped 3~uc SRO shows metallic behavior, whereas the 2~uc heterostructure is an insulator.

Figure 2(b) shows the effect of the STO capping layer on the Hall effect in 4~uc SRO film. Sohn {\it et al.}~\cite{sohn2018emergence} observed the hump structure in bare SRO ultra-thin films and argued that the ISB in ultra-thin films allows DMI which in turn induces the THE. We see that the 4~uc SRO film without an STO capping layer shows a strong hump structure. However, when the SRO films are capped with STO layers, the hump structure almost disappears and only the AHE part (the rectangular section of the hysteresis loop) remains (see Supplementary Material for details).

\begin{figure}[htbp]
	\includegraphics[width=0.46\textwidth]{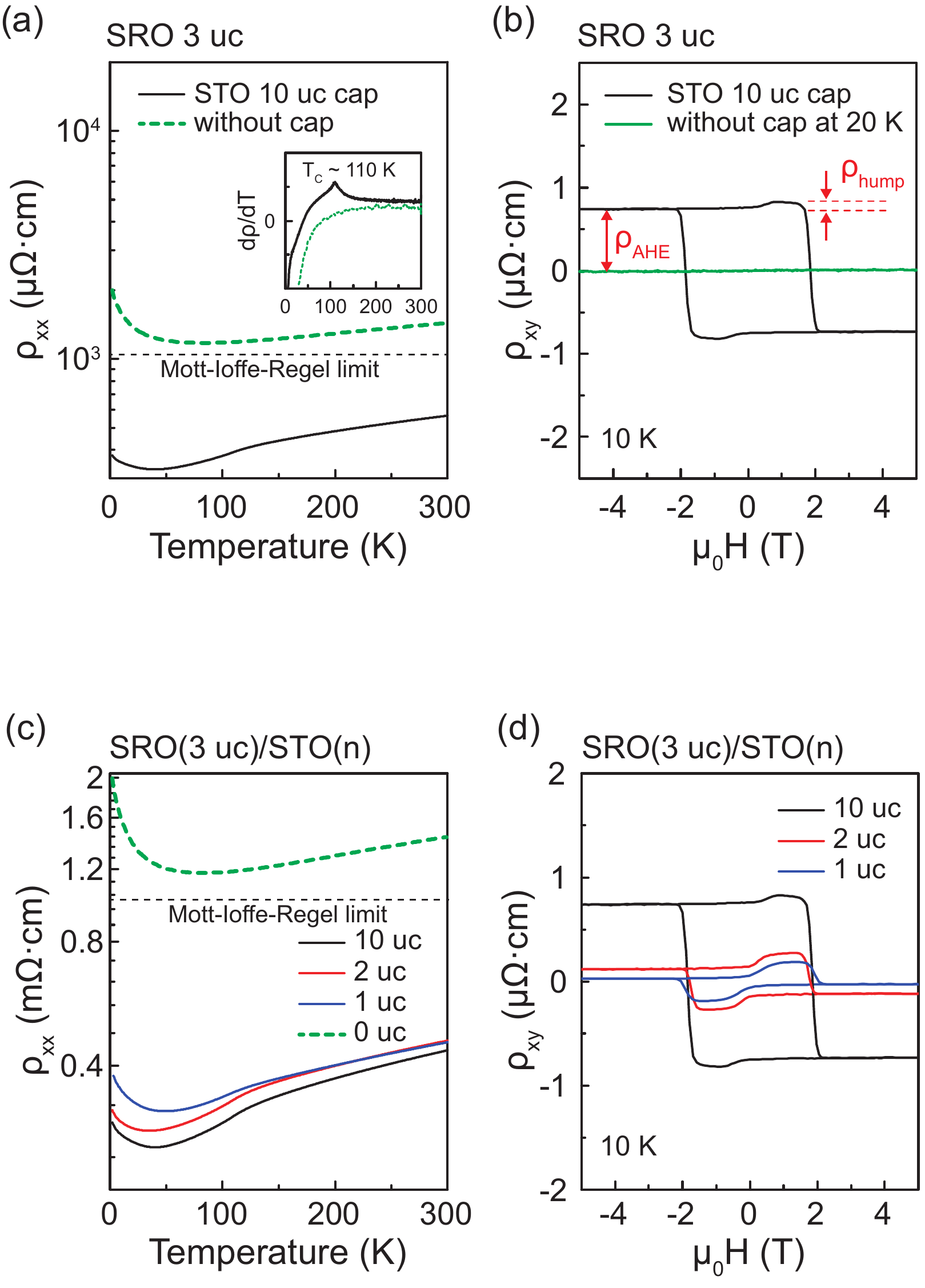}
	\caption{{\bf Effects of an STO capping layer on metallicity and Hall effects in 3~uc SRO.} ({\bf a}) Resistivity of 3~uc SRO thin films with and without a 10~uc STO capping layer. (Inset) Derivative of the resistivity for the two systems. Capping with an STO layer greatly reduces the resistivity and turns the film into a ferromagnet as evidenced by the kink in $d\rho/dT$. ({\bf b}) $\rho_{xy}$ of 3~uc SRO thin films with and without a 10~uc STO capping layer. The AHE hysteresis loop along with the hump structure shows that STO-capped 3~uc SRO is a ferromagnet, whereas no AHE signal is observed in the bare film. The anomalous Hall resistivity and size of hump structure are defined as $\rho_{\rm AHE}$ and $\rho_{\rm hump}$, respectively. ({\bf c}), ({\bf d}) STO thickness-dependent $\rho_{xx}$ and $\rho_{xy}$ of SRO(3~uc)/STO(n~uc).}
	\label{fig:3}
\end{figure}

It is well-known that the THE is enhanced in the ultra-thin regime~\cite{matsuno2016interface}. Hence, we can expect the hump structure to be stronger in thinner SRO films if the hump is related to the THE. As the STO-capped 3~uc SRO film is a ferromagnetic metal, as seen in Fig. 2(a), we decided to investigate the AHE and hump structure in 3~uc SRO films with various STO capping layer thicknesses. In Fig. 3(a), we compare resistivities of 3~uc SRO films with and without a 10~uc STO capping layer. 
The STO-capped 3~uc film is ferromagnetic (T$_c\approx 110~K$, as indicated by a kink in $d\rho/dT$) with a much reduced resistivity. This allows us to study the behavior of the Hall resistivity of STO-capped 3~uc SRO films.

Figure 3(b) shows the Hall measurement results. As discussed above, the 3~uc bare SRO film is nonmagnetic and we thus observe a vanishing Hall signal. On the other hand, clear AHE and hump structure appear in the STO-capped film. Following previous studies~\cite{matsuno2016interface, ohuchi2018electric}, we can explain the Hall resistivity of SRO films using three terms: $\rho_{xy} = \rho_{\rm OHE} + \rho_{\rm AHE} + \rho_{\rm hump}$ (the ordinary Hall effect (OHE), AHE, and hump structure, respectively). Here, the Hall data are presented without the OHE contribution. As shown in Fig. 3(b), we define $\rho_{\rm AHE}$ as the saturated $\rho_{xy}$ under a high magnetic field, and $\rho_{\rm hump}$ as the difference between the maximum $\rho_{xy}$ and $\rho_{\rm AHE}$.    

To shed more light on the capping effect, we measured $\rho_{xx}$ and $\rho_{xy}$ for various STO capping layer thicknesses. Figure 3(c) shows the $\rho_{xx}$ measurement results for various STO thicknesses. With only 1~uc of an STO capping layer, the metallicity is greatly enhanced as $\rho_{xx}$ decreases below the MIR limit. Thereafter, $\rho_{xx}$ slowly decreases with increasing STO capping layer thickness. Figure 3(d) shows the corresponding $\rho_{xy}$. It can be clearly seen that $\rho_{\rm AHE}$ increases with the thickness of the capping layer. On the other hand, $\rho_{\rm hump}$ decreases with the thickness. Note that the hump structure of the 3~uc SRO with 10~uc STO capping layer, though small, is more pronounced than that of the 10~uc STO-capped 4~uc SRO film. 

\begin{figure}[htbp]
	\includegraphics[width=0.46\textwidth]{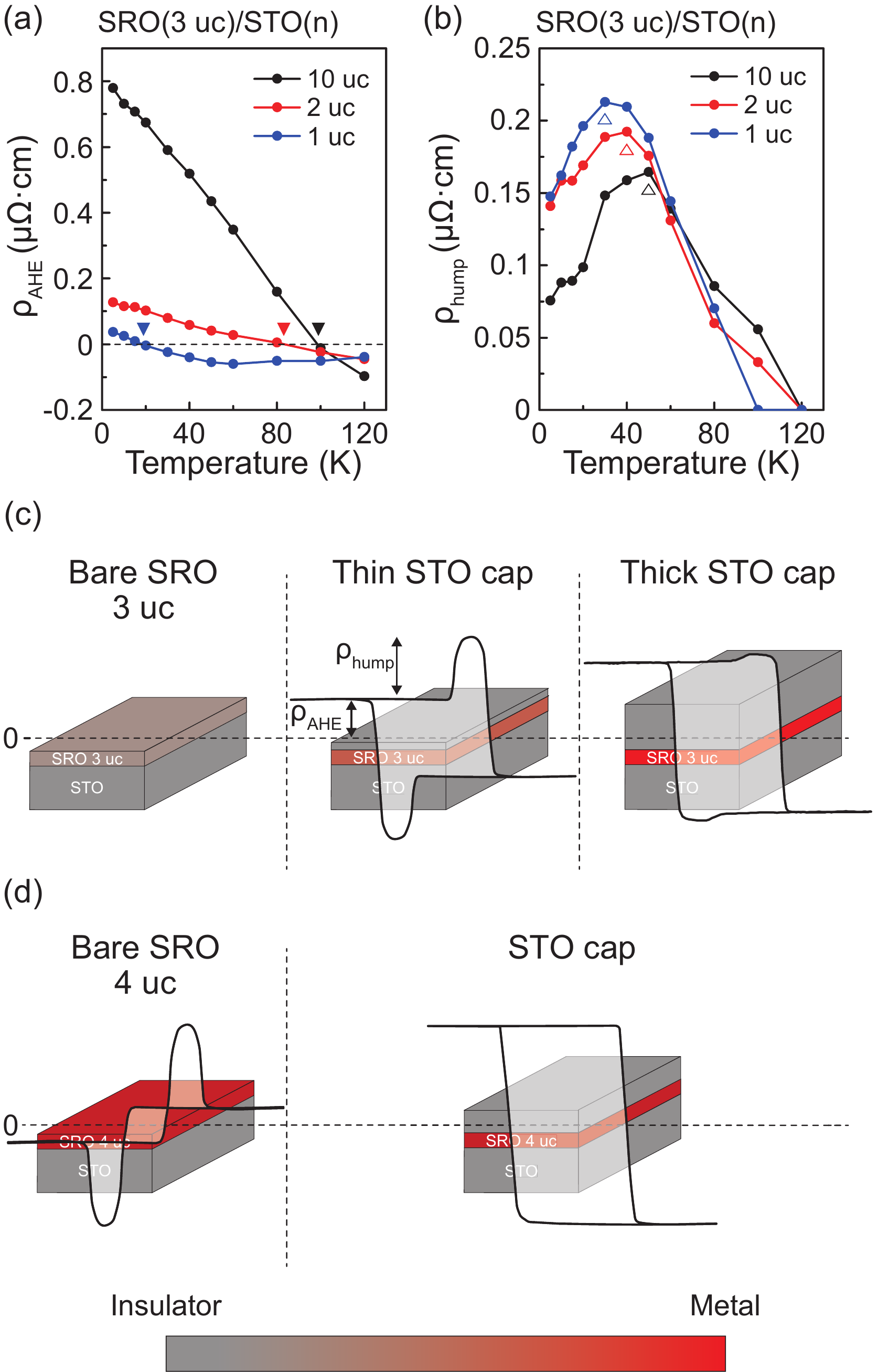}
	
	\caption{{\bf Temperature-dependent AHE and hump structure for various STO capping layer thicknesses.} Temperature-dependent ({\bf a}) $\rho_{\rm AHE}$ and (b) $\rho_{\rm hump}$ of SRO(3~uc)/STO(n~uc) films. Filled inverted triangles indicate the temperature for a sign change in $\rho_{\rm AHE}$, while open triangles mark the locations of maximum $\rho_{\rm hump}$. $\rho_{\rm hump}$ becomes smaller with a thicker STO capping layer and lower temperature. ({\bf c}) A schematic showing the effect of the STO capping layer on 3~uc SRO thin films. As the STO capping layer becomes thicker, $\rho_{\rm hump}$ decreases while $\rho_{\rm AHE}$ increases. ({\bf d}) A schematic showing the effect of the STO capping layer on 4~uc SRO thin films. The sign of AHE changes and the hump structure disappears.}
	\label{fig:4}
\end{figure}

We also performed temperature dependent studies on SRO(3~uc)/STO(n~uc). Figure 4(a) shows the temperature-dependent $\rho_{\rm AHE}$ in SRO(3~uc)/STO(n~uc). As the STO thickness increases, the overall $\rho_{\rm AHE}$ becomes larger, and the sign reversal temperature (filled inverted triangles) increases. It has previously been reported that the sign and magnitude of $\rho_{\rm AHE}$ can be controlled by the thickness of the film in the ultra-thin limit~\cite{sohn2019sign}. We find that $\rho_{\rm AHE}$ can be controlled not only by the thickness of the SRO, but also by the thickness of the capping layer. On the other hand, the maximum $\rho_{\rm hump}$, whose location is marked by empty triangles in Fig. 4(b), decreases with increasing STO capping layer thickness.

The effect of the STO capping layer on the SRO ultra-thin film is illustrated by schematics in Fig. 4(c) and (d). Bare SRO 3~uc does not exhibit AHE or hump structure due to its nonmagnetic nature. Adding an STO capping layer on the SRO 3~uc film turns the film into a ferromagnet with much lower resistivity and generates both the AHE and hump structure. As the thickness of the STO capping layer increases, $\rho_{\rm hump}$ decreases but $\rho_{\rm AHE}$ further increases. In the case of the 4~uc SRO film, the sign of $\rho_{\rm AHE}$ changes, and the hump structure disappears with an STO capping layer.

\begin{figure}[htbp]
	\includegraphics[width=0.46\textwidth]{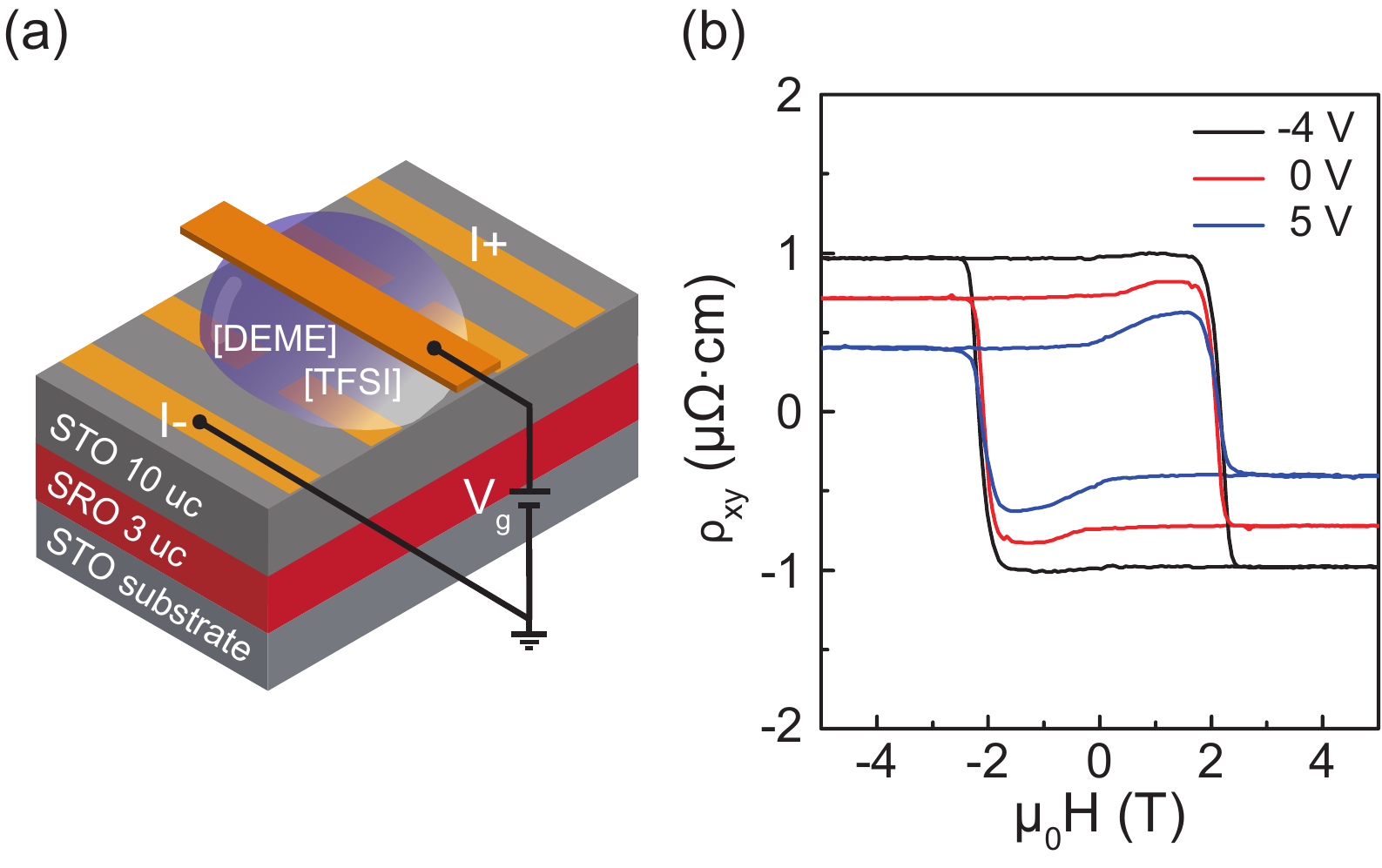}
	\caption{{\bf Hall effect measurement results from ionic liquid-gated SRO(3~uc)/STO(10~uc).} ({\bf a}) Schematic illustration of ionic liquid gating experiments. An electric field is applied on the SRO surface through the ionic liquid. ({\bf b}) Results for gate voltages of -4, 0, and 5~V. A positive (negative) electric voltage induces an inward (outward) electric field. $\rho_{\rm hump}$ is enhanced with a 5~V gate voltage, but disappears with a -4~V gate voltage.}
	\label{fig:5}
\end{figure}

Finally, we wish to demonstrate that we can control the AHE and hump structure via the ionic liquid gating method, which usually affects the surface/interface of heterostructures~\cite{cui2016manipulation}. We chose SRO(3~uc)/STO(10~uc) as the system since the STO capping layer is expected to prevent an irreversible chemical reaction in SRO~\cite{sharma2019ionic}. Figure 5(a) shows a schematic illustration of our ionic liquid gating experiment. We applied an electric field in the inward and outward directions, and measured the Hall resistivity as shown in Fig. 5(b). $\rho_{\rm AHE}$ decreases (increases) and $\rho_{\rm hump}$ increases (decreases) with positive (negative) gate voltage, which corresponds to a thinner (thicker) STO capping layer. Eventually, the hump structure almost disappears with a -4~V gate voltage.

It has been reported that DMI Hamiltonian, $H_{DMI} \equiv D_{ij} \cdot S_{i} \times S_{j}$, and magnetic skyrmions are controlled with ferroelectric proximity effect at the BaTiO$_3$/SrRuO$_3$ interface and THE in the SRO film varies~\cite{wang2018ferroelectrically}. In ionic liquid gating experiments, an electric field is applied between accumulated ions on the surface/interface and counter charges induced in the conductive layer~\cite{bisri2017endeavor}. In the positive bias case depicted in Fig. 5(a), a strong inward (relative to the substrate direction) electric field is applied to the surface of the SRO layer. We found that in several structural studies of perovskite oxides with ionic liquid gating, elongation of the out-of-plane lattice constant ($c$-axis lattice) has been reported with an inward electric field~\cite{dubuis2016oxygen,dong2017effect,walter2017ion} : (1) Coherent Bragg rod analysis studies of 5~uc La$_{1.96}$Sr$_{0.04}$CuO$_4$ film show that the displacement between the nearest site atoms increases (decreases) with an inward (outward) electric field~\cite{dubuis2016oxygen}, and (2) XRD studies report elongation of the $c$-axis lattice with an inward electric field, regardless of the majority carriers~\cite{dong2017effect,walter2017ion}. Thus, we speculate that the elongation of the out-of-plane lattice constant and interface distortion induced by the gate voltage can cause the variation in DMI and $\rho_{\rm hump}$. 

Despite intensive recent studies, the origin of the hump structure observed in the Hall effect data from bare SRO film is still under debate~\cite{sohn2018emergence, sohn2020hump, kan2018alternative,kimbell2020two,wang2020controllable,wu2018artificial,wu2020berry,wysocki2020electronic,wysocki2020validity}. It has been suggested that atomic rumpling of SRO ultra-thin film can induce sufficient DMI to form magnetic skyrmions~\cite{sohn2018emergence}. An alternative interpretation of the hump structure has also been proposed, in which a combination of different AHEs that arise due to thickness inhomogeneity can lead to hump structures in the Hall data~\cite{kan2018alternative,kimbell2020two,wang2020controllable,wu2018artificial,wu2020berry}. We believe the behavior of $\rho_{hump}$ in our films can be better explained within the former view because the atomic structure at the interface is directly affected by capping layers and gate voltages, However, we also acknowledge that further studies are needed to pin down the true cause of the hump structure.

In summary, both the AHE and hump structure in the Hall resistivity of SRO ultra-thin films can be modulated by capping the film with an STO layer and gating with an ionic liquid. The AHE and hump structure change significantly when STO layers are capped on SRO ultra-thin films. We controlled the electric field on the SRO/STO interface by using the ionic liquid gating method and found that the AHE and hump structure can be tuned with the gate voltage. We attribute the large modulation of the AHE and hump structure to an electric field tuned inversion symmetry, which is related to DMI.

\section*{Supplementary Material}
See supplementary material for the details of experimental data.

\acknowledgments
The authors wish to thank J. R. Kim and Y. Ishida for fruitful discussions. This work is supported by the Institute for Basic science in Korea (Grant No. IBS-R009-G2). PPMS measurements are supported by the National Center for Inter-University Research Facilities (NCIRF) at Seoul National University in Korea. The work at Yonsei University is supported by the National Research Foundation of Korea (NRF) Grants (NRF-2017R1A5A1014862 (SRC program: vdWMRC center) and NRF-2019R1A2C2002601).

\section*{data availability}
The data that support the findings of this study are available from the corresponding author upon reasonable request.

%

\end{document}